\documentclass[12pt]{article}

\usepackage{graphics,amssymb,epsfig,float}
\usepackage[usenames,dvips]{color}
\usepackage{graphicx}% Include figure files
\usepackage{epsfig}% Include figure files
\usepackage{rotating}% Include figure files
\usepackage{dcolumn}% Align table columns on decimal point
\usepackage{bm}% bold math
\usepackage{cite}
\usepackage{amsmath}
\usepackage{setspace}% double spacing

\def\lsim{\;\raise0.3ex\hbox{$<$\kern-0.75em\raise-1.1ex\hbox{$\sim$}}\;}
\def\gsim{\;\raise0.3ex\hbox{$>$\kern-0.75em\raise-1.1ex\hbox{$\sim$}}\;}
\def\beq{\begin{equation}}   \def\eeq{\end{equation}}
\def\ba{\begin{array}}       \def\ea{\end{array}}
\def\bea{\begin{eqnarray}}   \def\eea{\end{eqnarray}}
\def\nn{\nonumber}

\begin{document}

\begin{titlepage}
\begin{flushright}
LPT Orsay 13-08\\
PCCF RI 13-01
\end{flushright}

\begin{center}
\vspace{1cm}
{\Large\bf LHC constraints on $M_{1/2}$ and $m_0$ in the
semi-constrained NMSSM}
\vspace{2cm}

{\bf{Debottam Das$^1$, Ulrich Ellwanger$^1$ and Ana M. Teixeira$^2$}}
\vspace{1cm}\\
\it $^1$  Laboratoire de Physique Th\'eorique, UMR 8627, CNRS and
Universit\'e de Paris--Sud,\\
\it B\^at. 210, 91405 Orsay, France \\
\it $^2$ Laboratoire de Physique Corpusculaire, CNRS/IN2P3 -- UMR
6533,\\
\it Campus des C\'ezeaux, 24 Av. des Landais, F-63171 Aubi\`ere Cedex,
France\\

\end{center}

\vspace{1cm}

\begin{abstract}
Constraints from searches for squarks and gluinos at the LHC at
$\sqrt{s}$=8~TeV are applied to the parameter space of the NMSSM with
universal squark/slepton and gaugino masses at the GUT scale, but
allowing for non-universal soft Higgs mass parameters (the sNMSSM). We
confine ourselves to regions of the parameter space compatible with a
125~GeV Higgs boson with diphoton signal rates at least as large as the
Standard Model ones, and a dark matter candidate compatible with WMAP
and XENON100 constraints. Following the simulation of numerous points in the
$m_0-M_{1/2}$ plane, we compare the constraints on the sNMSSM from 3-5
jets + missing $E_T$ channels as well as from multijet + missing $E_T$
channels with the corresponding cMSSM constraints. Due to the longer
squark decay cascades, lower bounds on $M_{1/2}$ are alleviated by up to
50~GeV. For heavy squarks at large $m_0$, the dominant constraints
originate from multijet + missing $E_T$ channels due to gluino decays
via stop pairs.
\end{abstract}

\end{titlepage}

\newpage
\section{Introduction}

One of the most important tasks of the LHC -- besides the
quest for the Higgs boson -- is the search for new elementary particles
like those predicted in supersymmetric (SUSY) extensions of the
Standard Model (SM). So far the search for such SUSY particles
(sparticles) has not been successful; the absence of corresponding
signal events can be interpreted as lower bounds on sparticle masses
(see \cite{ATLAS-CONF-2012-103,ATLAS-CONF-2012-104,ATLAS-CONF-2012-105,
ATLAS-CONF-2012-109,ATLAS-CONF-2013-001,ATLAS-CONF-2012-151} for recent
ATLAS publications of results at $\sqrt{s}$=8~TeV,
\cite{CMS-PAS-SUS-12-016,CMS-PAS-SUS-12-017,
CMS-PAS-SUS-12-018,CMS-PAS-SUS-12-023,CMS-PAS-SUS-12-028} for recent CMS
publications of results at $\sqrt{s}$=8~TeV, \cite{deJong:2012zt} and
the web pages \cite{ATLASweb} and \cite{CMSweb} for summaries of
searches for sparticles by the ATLAS and CMS collaboration). Clearly
these lower bounds on sparticle masses are not model independent, since
they depend on the sparticle couplings and decay cascades, and hence on a
large number of unknown parameters.

The large number of unknown parameters of SUSY extensions
of the SM is greatly reduced if one assumes universal soft SUSY
breaking terms at the Grand Unification (GUT) scale, which is also
theoretically appealing. Such models are denoted as ``constrained'', and
this is the case of
the cMSSM (constrained Minimal Supersymmetric extension of the SM).
Since the various sparticle masses and couplings are now strongly
correlated, constrained models often serve as useful benchmark
scenarios. The bounds on sparticle masses can then be represented as
bounds in the $m_0-M_{1/2}$ plane, where $m_0$ denotes the universal
squark and slepton masses and $M_{1/2}$ the universal gaugino masses at
the GUT scale. Frequently the cMSSM with $\tan\beta = 10$
and $A_0=0$ is used to this end ($\tan\beta$ being the ratio of the
two Higgs vevs $\left<H_u\right>/\left<H_d\right>$, and $A_0$ denoting
the universal soft SUSY breaking trilinear couplings at the GUT scale).

However, most of the considered parameter space of the cMSSM with
$\tan\beta = 10$ and $A_0=0$ is neither consistent with the observation
of a SM like Higgs boson near 125~GeV \cite{:2012gk,:2012gu}, nor with
the dark matter relic density as determined by the WMAP experiment
\cite{Komatsu:2010fb}. In recent publications
\cite{Bechtle:2012zk,Arbey:2012na,Fowlie:2012im,CahillRowley:2012cb,
Beskidt:2012sk,CahillRowley:2012kx,Mahmoudi:2012eh}, the LHC bounds on
sparticle masses have been applied to the cMSSM (or variants thereof as
the NUHM with non-universal soft Higgs masses at the GUT scale), but
with parameters consistent with a SM like Higgs boson near 125~GeV,
and/or dark matter consistent with WMAP bounds 
on the relic density and XENON100 limits
\cite{Aprile:2012nq} on the dark matter direct detection cross section.
Generally, one finds that the bounds on sparticle masses
obtained within variants of the cMSSM or in the NUHM are similar to the
cMSSM with $\tan\beta = 10$ and $A_0=0$.

The MSSM is not the only possible supersymmetric extension of the SM.
The simplest super\-symmetric extension of the Standard Model with a
scale invariant superpotential, i.e. where the soft SUSY breaking terms
are the only dimensionful parameters, is the Next-to-Minimal
Supersymmetric Standard Model (NMSSM) \cite{Ellwanger:2009dp}.  A
supersymmetric Higgs mass term $\mu$, as required in the MSSM, is
generated dynamically by a vacuum expectation value (vev) of a gauge
singlet (super-)field $S$, and is naturally of the order of the
SUSY breaking scale. The attractive features of the MSSM are
preserved, like a solution of the hierarchy problem, the unification of
the running gauge coupling constants at a Grand Unification scale, and a
dark matter candidate in the form of a stable lightest SUSY
particle~(LSP).

The additional coupling $\lambda$ of the Higgs bosons in the NMSSM makes
it much easier to accommodate a SM like Higgs boson near 125~GeV
\cite{Arvanitaki:2011ck,Hall:2011aa,Ellwanger:2011aa,
Gunion:2012zd,
King:2012is,Kang:2012sy,Cao:2012fz,Vasquez:2012hn,Ellwanger:2012ke,
Jeong:2012ma,Benbrik:2012rm,Gunion:2012gc,Cao:2012yn,Cheng:2012pe,
SchmidtHoberg:2012yy,Kang:2012bv,Belanger:2012sd,Agashe:2012zq,
Belanger:2012tt, Heng:2012at,Choi:2012he,King:2012tr,Gherghetta:2012gb,
Kang:2013rj}
and alleviates the corresponding ``little fine-tuning problem'' of the
cMSSM since lighter scalar top quarks (top squarks) are possible
\cite{Arvanitaki:2011ck,Hall:2011aa,Ellwanger:2011aa,King:2012is,Kang:2012sy,
Cao:2012fz,Ellwanger:2012ke,Jeong:2012ma,Cao:2012yn,Cheng:2012pe,
SchmidtHoberg:2012yy,Agashe:2012zq,Heng:2012at,King:2012tr,Gherghetta:2012gb,
Kang:2012bv,Ellwanger:2011mu}. The fermionic component of the gauge
singlet superfield $\hat{S}$ extends the neutralino sector of the MSSM,
and this
can lead to more complicated sparticle cascade
decays~\cite{Ellwanger:1997jj, Ellwanger:1998vi,Barger:2006kt,
Das:2012rr,Vasquez:2012hn,Dreiner:2012ec}. 

Like in the MSSM, one can consider constrained versions of the NMSSM
with universal soft SUSY breaking terms at the GUT scale. A
SM like Higgs boson near 125~GeV can easily be obtained within the
semi-constrained NMSSM in which, similar to the NUHM, the soft SUSY
breaking Higgs mass terms (and the trilinear couplings involving the
singlet $S$) are allowed to deviate from the soft SUSY breaking terms
involving squarks or sleptons. In several recent publications the
parameter space of the semi-constrained NMSSM compatible with a SM like
Higgs boson near 125~GeV (and possibly an enhanced diphoton signal rate)
has been discussed \cite{Gunion:2012zd,Ellwanger:2012ke,
Gunion:2012gc,Gunion:2011hs,Belanger:2012tt}. The LHC constraints from
negative squark and gluino searches had to be estimated or had
been left aside, unless trivially satisfied due to
very heavy squarks and/or gluinos. LHC constraints from the runs at
$\sqrt{s}=7$~TeV, using the razor variables at CMS
\cite{Chatrchyan:2011ek}, have been studied within a more restricted
version of the semi-constrained NMSSM allowing {\it only} the soft SUSY
breaking singlet mass term to deviate from $m_0$ in
\cite{Kowalska:2012gs}. In this case the regions in the parameter space
corresponding to large $\lambda$, low $m_0$ and $M_{1/2}$, which are
interesting from the point of view of a 125~GeV Higgs with low
fine-tuning, are not viable.

Hence it becomes interesting and important to re-analyse the LHC
constraints on sparticle masses within the semi-constrained NMSSM, which
is the purpose of the present paper. We focus on the most constraining
squark and gluino search channels analysed by the
ATLAS collaboration for the $\sqrt{s}=8$~TeV run: Searches for final 
states with jets and missing 
transverse momentum \cite{ATLAS-CONF-2012-109}, final states with
large jet multiplicities \cite{ATLAS-CONF-2012-103}, and with an
isolated lepton \cite{ATLAS-CONF-2012-104}.

We confine ourselves to phenomenologically acceptable regions of the
sNMSSM parameter space with a Higgs boson
near 125~GeV, a diphoton signal rate near or above its SM value, a
dark matter relic density in agreement with WMAP constraints, and a 
dark matter direct
detection cross section compatible with XENON100 constraints. The latter
constrain the neutralino sector (the mass and the couplings of the
lightest SUSY particle, the LSP) which has some impact on the sparticle
decay cascades.

In the NMSSM, sparticle decay cascades can differ from the MSSM for
various reasons:
\begin{itemize}
\item The higgsino-like neutralinos and charginos can be lighter than in
most realistic scenarios within the MSSM (with a Higgs boson near
125~GeV and a dark matter relic density compatible with WMAP
constraints). The additional singlet-like neutralino (singlino) can
mix with the MSSM like neutralinos, implying more complicate sparticle
cascade decays~\cite{Ellwanger:1997jj,Ellwanger:1998vi,Barger:2006kt,
Das:2012rr,Vasquez:2012hn,Dreiner:2012ec}. 
\item The top squarks are typically lighter than in realistic scenarios
within the cMSSM (with a Higgs boson near 125~GeV), implying
gluino/squark cascade decays via top squarks which lead to multijet
events, reducing the missing transverse momentum and the
average transverse momenta of jets.
\end{itemize}

Hence it is not clear to which extent the bounds in the $m_0-M_{1/2}$ plane
obtained by ATLAS for the cMSSM with $\tan\beta=10$ and $A_0=0$ are
applicable to the semi-constrained NMSSM; the results of our present
study allow to answer
this question quantitatively.

In the next section, we briefly review the semi-constrained NMSSM and
discuss our choice for points in the $m_0-M_{1/2}$ plane. In
Section~3 we describe the tools used for the Monte Carlo studies and
list the applied LHC constraints. In Section~4 we present the resulting
bounds in the $m_0-M_{1/2}$ and $M_{\text{squark}} - 
M_{\text{gluino}}$ planes, discuss the origin of the differences in the bounds within
the semi-constrained NMSSM with respect to the cMSSM, and summarise 
our conclusions.

\section{The semi-constrained NMSSM}

The NMSSM differs from the MSSM due to the presence of the gauge singlet
superfield $\hat S$. In the simplest realisation of the NMSSM, the 
$\mu \hat H_u \hat H_d$ Higgs mass
term in the MSSM superpotential $W_\text{MSSM}$ 
is replaced by the coupling $\lambda$ of $\hat S$ to $\hat H_u$ and
$\hat H_d$, and a self-coupling $\kappa \hat S^3$.  Hence, in this version
the superpotential $W_\text{NMSSM}$ is scale invariant, and given by:
\beq\label{eq:1}
W_\text{NMSSM} = \lambda \hat S \hat H_u\cdot \hat H_d + \frac{\kappa}{3} 
\hat S^3 +\dots\; ,
\eeq
where the dots denote the Yukawa couplings of $\hat H_u$ and $\hat H_d$
to the quarks and leptons as in the MSSM. Once the scalar component of
$\hat S$ develops a vev $s$, the first term in $W_\text{NMSSM}$ generates an
effective $\mu$-term with
\beq\label{eq:2}
\mu_\mathrm{eff}=\lambda\, s\; .
\eeq

The soft SUSY breaking terms consist of mass terms for the Higgs bosons
$H_u$, $H_d$, $S$, squarks
$\tilde{q_i} \equiv (\tilde{u_i}_L, \tilde{d_i}_L$), $\tilde{u_i}_R^c$,
$\tilde{d_i}_R^c$ and sleptons $\tilde{\ell_i} \equiv (\tilde{\nu_i}_L,
\tilde{e_i}_L$) and $\tilde{e_i}_R^c$ 
(where $i=1\dots 3$ is a generation index):
\bea
-{\cal L}_\mathrm{0} &=&
m_{H_u}^2 | H_u |^2 + m_{H_d}^2 | H_d |^2 + 
m_{S}^2 | S |^2 +m_{\tilde{q_i}}^2|\tilde{q_i}|^2 
+ m_{\tilde{u_i}}^2|\tilde{u_i}_R^c|^2
+m_{\tilde{d_i}}^2|\tilde{d_i}_R^c|^2\nn \\
&& +m_{\tilde{\ell_i}}^2|\tilde{\ell_i}|^2
+m_{\tilde{e_i}}^2|\tilde{e_i}_R^c|^2\; ,
\label{eq:3}
\eea
trilinear interactions involving the third generation squarks, sleptons
and the Higgs fields (neglecting the Yukawa couplings of the first two
generations):
\bea
-{\cal L}_\mathrm{3}&=& 
\Bigl( h_t A_t\, Q\cdot H_u \: \tilde{u_3}_R^c +
h_b  A_b\, H_d \cdot Q \: \tilde{d_3}_R^c +
h_\tau A_\tau \,H_d\cdot L \: \tilde{e_3}_R^c  \nn \\
&& +\,  \lambda A_\lambda\, H_u  \cdot H_d \,S +  \frac{1}{3} \kappa 
A_\kappa\,  S^3 \Bigl)+ \, \mathrm{h.c.}\;,
\label{eq:4}
\eea
and mass terms for the gauginos $\tilde{B}$ (bino), $\tilde{W}^a$
(winos) and $\widetilde{G}^a$ (gluinos):
 \beq\label{eq:5}
-{\cal L}_\mathrm{1/2}= \frac{1}{2} \bigg[ 
 M_1 \widetilde{B}  \widetilde{B}
\!+\!M_2 \sum_{a=1}^3 \widetilde{W}^a \widetilde{W}_a 
\!+\!M_3 \sum_{a=1}^8 \widetilde{G}^a \widetilde{G}_a   
\bigg]+ \mathrm{h.c.}\;.
\eeq

The neutral CP-even Higgs sector contains 3 states $H_i$, which are
mixtures of the CP-even components of the superfields $\hat H_u$, $\hat
H_d$ and $\hat S$. Their masses are described by a $3 \times 3$ mass
matrix ${\cal M}^2_{H\,ij}$. The neutral CP-odd Higgs sector contains 2
physical states $A_i$, whose masses are described by a $2 \times 2$ mass
matrix ${\cal M}^2_{A\,ij}$. In the neutralino sector we have 5 states
${\chi}^0_i$, which are mixtures of the bino $\widetilde{B}$, the neutral
wino $\widetilde{W}^3$, the neutral higgsinos from the superfields $\hat
H_u$ and $\hat H_d$, and the singlino from the superfield $\hat S$.
Their masses are described by a $5 \times 5$ mass matrix ${\cal
M}_{\chi^0\,ij}$. Expressions for the mass matrices  -- after $H_u$,
$H_d$ and $S$ have developed vevs $v_u$, $v_d$ and $s$, and including the
dominant radiative corrections -- can be found
in~\cite{Ellwanger:2009dp} and will not be repeated here.

As compared to two independent parameters in the Higgs sector of the
MSSM at tree level (often chosen as $\tan \beta$ and $M_A$), the Higgs
sector of the NMSSM contains six para\-me\-ters
\beq \label{eq:6}
\lambda\ , \ \kappa\ , \ A_{\lambda} \ , \ A_{\kappa}, \ \tan\beta\equiv
v_u/v_d\ ,\ \mu_\mathrm{eff}\; ;
\eeq
then the soft SUSY breaking mass terms for the Higgs bosons $m_{H_u}^2$,
$m_{H_d}^2$ and $m_{S}^2$ are determined implicitely by $M_Z$,
$\tan\beta$ and $\mu_\mathrm{eff}$.

In constrained versions of the NMSSM (as in the constrained MSSM) one
assumes that the soft SUSY breaking terms involving gauginos, squarks
and
sleptons are universal at the GUT scale:
\beq \label{eq:7}
M_1 = M_2 = M_3 \equiv M_{1/2}\; ,
\eeq
\beq \label{eq:8}
m_{\tilde{q}_i}^2= m_{\tilde{u}_i}^2=m_{\tilde{d}_i}^2=
m_{\tilde{\ell}_i}^2=m_{\tilde{e}_i}^2\equiv m_0^2\; ,
\eeq
\beq \label{eq:9}
A_t = A_b = A_\tau \equiv A_0\; .
\eeq

In the semi-constrained NMSSM considered here, one allows
the Higgs sector to play a special r\^ole: the Higgs soft mass terms
$m_{H_u}^2$, $m_{H_d}^2$ and $m_{S}^2$ are allowed to differ from
$m_0^2$ (and determined implicitely as noted above), and the trilinear
couplings $A_{\lambda}$, $A_{\kappa}$ can differ from $A_0$. Hence the
complete parameter space is characterised by
\beq \label{eq:10}
\lambda\ , \ \kappa\ , \ \tan\beta\ ,\
\mu_\mathrm{eff}\ , \ A_{\lambda} \ , \ A_{\kappa} \ , \ A_0 \ , \
M_{1/2}\ ,  \ m_0\; ,
\eeq
where the latter five parameters are taken at the GUT scale.

Subsequently we are interested in regions of the parameter space with
large NMSSM-specific contributions to the SM-like Higgs mass, i.e. large
values of $\lambda$ (and $\kappa$) and low values of $\tan\beta$, which
lead naturally to a SM-like Higgs boson $H_2$ in the 125~GeV
range~\cite{Arvanitaki:2011ck,Hall:2011aa,Ellwanger:2011aa,Gunion:2012zd,
King:2012is,Kang:2012sy,Cao:2012fz,Vasquez:2012hn,Ellwanger:2012ke,
Jeong:2012ma,Benbrik:2012rm,Gunion:2012gc,Cao:2012yn,Cheng:2012pe,
SchmidtHoberg:2012yy,Kang:2012bv,Belanger:2012sd,Agashe:2012zq,
Belanger:2012tt, Heng:2012at,Choi:2012he,King:2012tr,Gherghetta:2012gb,
Kang:2013rj}. 
We impose constraints from
LEP~\cite{Schael:2006cr}
on the lighter
mostly singlet-like Higgs boson $H_1$, which still allow for a $H_1$ mass below
114~GeV if its coupling to the $Z$~boson is reduced. For
$H_2$ we require  $124\ \mathrm{GeV} < M_{H_2} < 127\ \mathrm{GeV}$, 
$\sigma_\text{obs}^{\gamma\gamma}(H_2)/ \sigma_\text{SM}^{\gamma\gamma} > 1$ and
$\sigma_\text{obs}^{ZZ}(H_2)/ \sigma_\text{SM}^{ZZ} \sim 1$ in order to comply
with the observations at the LHC.

We have implemented these constraints into a modified version of the
public code \hbox{NMSPEC}~\cite{Ellwanger:2006rn} inside
NMSSMTools~\cite{Ellwanger:2004xm,Ellwanger:2005dv}. (In the Higgs
sector we have used two-loop radiative corrections
from~\cite{Degrassi:2009yq}, and for the top quark pole mass we have
taken
$m_\text{top}=173.1$~GeV.) The constraints from $B$-physics are those of
the version 3.2.0 of NMSSMTools, which are easily satisfied for 
the regime
$\tan\beta<3$ relevant here. The dark matter relic density and
direct detection cross section of the LSP $\chi^0_1$ (the lightest
neutralino) are computed by
MicrOmegas~\cite{Belanger:2005kh,Belanger:2006is,Belanger:2008sj}
implemented in NMSSMTools. However, due to the low values of
$\tan\beta$, the SUSY contribution to the anomalous
magnetic moment of the muon $\Delta a_\mu$ is not large enough to resolve the
discrepancy between the SM and its measured value.

Leaving aside $\Delta a_\mu$, many regions in the space of the parameters
in \eqref{eq:10} satisfy all the above conditions. Hence we proceed as
follows: We start with a ``lattice'' in the $m_0-M_{1/2}$ plane, i.e.
numerous fixed values for $m_0$ and $M_{1/2}$. For each fixed ($m_0$,
$M_{1/2}$) we choose the remaining parameters such that not only the
above phenomenological constraints are satisfied, but also in such a way
 that the
lighter top squark mass and $\mu_\mathrm{eff}$ are relatively small.

Light stop quarks and low $\mu_\mathrm{eff}$ minimise the fine-tuning
\cite{Arvanitaki:2011ck,Hall:2011aa,Ellwanger:2011aa,King:2012is,Kang:2012sy,
Cao:2012fz,Ellwanger:2012ke,Jeong:2012ma,Cao:2012yn,Cheng:2012pe,
SchmidtHoberg:2012yy,Agashe:2012zq,Heng:2012at,King:2012tr,Gherghetta:2012gb,
Kang:2012bv,Ellwanger:2011mu}. On the other hand,
present constraints from searches for these sparticles should be
satisfied. First, since gluinos with masses below $\sim 1$~TeV are
excluded, constraints from gluino mediated stop production
\cite{CMS-PAS-SUS-12-016,Aad:2012pq, ATLAS:2012ah,ATLAS-CONF-2012-151}
turn out to be satisfied. Constraints from direct pair production of top
squarks~$\tilde{t}$ \cite{:2012si,:2012ar, :2012yr,Aad:2012uu,CMS-PAS-SUS-11-022,
ATLAS-CONF-2013-001} require $m_{\tilde{t}_1} \gsim 400$~GeV for
LSP masses of $\sim 80$~GeV, as found below (note that slightly stronger
bounds assume branching ratios and neutralino/chargino masses within
simplified models which are not valid here; see
\cite{Bi:2012jv} for proposals for search strategies for light stops
within the general NMSSM). In addition, we require $\mu_\mathrm{eff}
\gsim 120$~GeV so that the lighter chargino masses and
chargino-neutralino mass splittings comply with present constraints.

For each such point on a lattice in the $m_0-M_{1/2}$ plane, we perform
Monte Carlo simulations ($\sim 10^4$ events), apply the cuts
described in the next section, and compare the resulting signal event
numbers to present constraints. This allows to identify viable
regions (up to error bars) in the $m_0-M_{1/2}$ and $M_{\text{squark}} -
M_{\text{gluino}}$ planes within the semi-constrained NMSSM.

\section{Monte Carlo simulation, search channels and verification}

For the calculation of the matrix elements we use MadGraph/MadEvent~5
\cite{Alwall:2011uj}, which includes Pythia~6.4 \cite{Sjostrand:2006za}
for showering and hadronisation. Matching of the differential jet cross
sections is performed according to the prescriptions in
\cite{Alwall:2008qv}. The sparticle branching ratios are obtained with
the help of the code NMSDECAY \cite{Das:2011dg} (based on SDECAY
\cite{Muhlleitner:2003vg}), and are passed to Pythia.

The output is given in StdHEP-format to the fast detector simulation
Delphes \cite{Ovyn:2009tx}. Inside Delphes, the anti-k(t) jet
reconstruction algorithm \cite{Cacciari:2008gp} is used, with the jet
reconstruction performed by FastJet \cite{Cacciari:2005hq}.

The sparticle (squark and gluino) production cross sections are obtained
by Prospino at next-to-leading order (NLO) \cite{Beenakker:1996ch,
Beenakker:1996ed,Beenakker:1999xh}. The resummation of soft gluon
emission is taken into account in the form of a correction factor
estimated from \cite{Beenakker:2011fu}, and the theoretical
uncertainties from scale and PDF choices are obtained from
\cite{Beenakker:2011fu,Kramer:2012bx}.

To the output from Delphes we apply cuts on final states with jets and
missing transverse momentum from searches for supersymmetry at
$\sqrt{s}=8$~TeV with an integrated luminosity of 5.8~fb$^{-1}$ by the
ATLAS collaboration \cite{ATLAS-CONF-2012-109}, which give 
at present the strongest
constraints in the $m_0-M_{1/2}$ plane in the cMSSM, as well
as cuts on final states with large jet multiplicities from
\cite{ATLAS-CONF-2012-103} and one isolated lepton from
\cite{ATLAS-CONF-2012-104} which could, a priori, be relevant for the
NMSSM.

In Table~1 we summarise the cuts corresponding to the search channels
which lead to the most stringent constraints in the $m_0-M_{1/2}$ plane
in the semi-constrained NMSSM (depending on $m_0$ and $M_{1/2}$), and
the 95\% confidence level (CL) upper limits (UL) on the number $N_{SE}$
of signal events beyond the expected background in the corresponding
channels, for an integrated luminosity of 5.8~fb$^{-1}$. (More details on
the event selections can be found in \cite{ATLAS-CONF-2012-103,
ATLAS-CONF-2012-109}; bounds from the searches including one isolated
lepton did not lead to stronger constraints.) Here
$m_{\mathrm{eff}}$(Nj) is the scalar sum of the transverse momenta of
$E_T^{\mathrm{miss}}$ together with the leading N jets,
$m_{\mathrm{eff}}$(incl.) the scalar sum of the transverse momenta of
$E_T^{\mathrm{miss}}$ together with all jets with $p_T > 40$~GeV, and
$H_T$ the scalar sum of the transverse momenta of all jets with $p_T >
40$~GeV without $E_T^{\mathrm{miss}}$.

\begin{table}[t!]
\begin{center}
\begin{tabular}{|c|c|c|c|c|c|} \hline
Channel & C-tight \cite{ATLAS-CONF-2012-109} &
D-tight \cite{ATLAS-CONF-2012-109} &
E-tight \cite{ATLAS-CONF-2012-109} &
9j55 \cite{ATLAS-CONF-2012-103} &
8j80 \cite{ATLAS-CONF-2012-103}  \\\hline
$N_{jet}$ & 4 & 5 & 6 & $\ge 9$ & $\ge 8$\\\hline
$E_T^{\mathrm{miss}}>$  & 160 & 160 & 160 & &\\\hline
$E_T^{\mathrm{miss}}/\sqrt{H_T}>$ & & & & 4
& 4 \\\hline
$p_T(j_1)>$ & 130 & 130 & 130 & 55  & 80 \\\hline
$p_T(j_2\dots j_{N_{jet}})>$ & 60 & 60 & 60 & 55 &80  \\\hline
$m_{\mathrm{eff}}$(incl.)$>$ & 1900 & 1700 & 1400 & & \\\hline
$E_T^{\mathrm{miss}}/m_{\mathrm{eff}}$(Nj)$>$ & 0.25 & 0.15 & 0.15& &
\\\hline
 95\% CL UL on $N_{SE}$ & 3.3 & 6.0 & 9.3 & 5.4 & 4.0 \\\hline
\end{tabular}
\end{center}
\caption{Cuts and 95\% CL upper limits on the number $N_{SE}$ of signal
events at $\sqrt{s}=8$~TeV and 5.8~fb$^{-1}$ integrated luminosity for the search
channels leading to the most stringent constraints in the $m_0-M_{1/2}$
plane in the cMSSM or the semi-constrained NMSSM.
($E_T^{\mathrm{miss}}$, $p_T$, $H_T$ and $m_{\mathrm{eff}}$ in GeV.)}
\label{tab:1}
\end{table}

We first verified the validity of our simulations in the framework of the cMSSM with
$\tan\beta = 10$ and $A_0=0$: For points in the $m_0-M_{1/2}$ plane
along the 95\% CL exclusion line in \cite{ATLAS-CONF-2012-109} we
determined the number of signal events in all search channels,  divided
them by the corresponding upper limits given in
\cite{ATLAS-CONF-2012-109}, and computed a ratio $R$ from the most
constraining search channel (giving the largest value for $R$). If our
simulations would coincide exactly with those in
\cite{ATLAS-CONF-2012-109}, we would obtain $R=1$. The
resulting ratio $R$ is shown as function of $m_0$ in Figure~1, where we
also indicate the most constraining search channel by colours: black
(full) for C-tight, blue (dashed) for D-tight and green (dotted) for
E-tight. The most constraining search channels, depending on $m_0$,
coincide with the information given in \cite{ATLAS-CONF-2012-109}.

\begin{figure}[h!]
\begin{center}
\epsfig{file=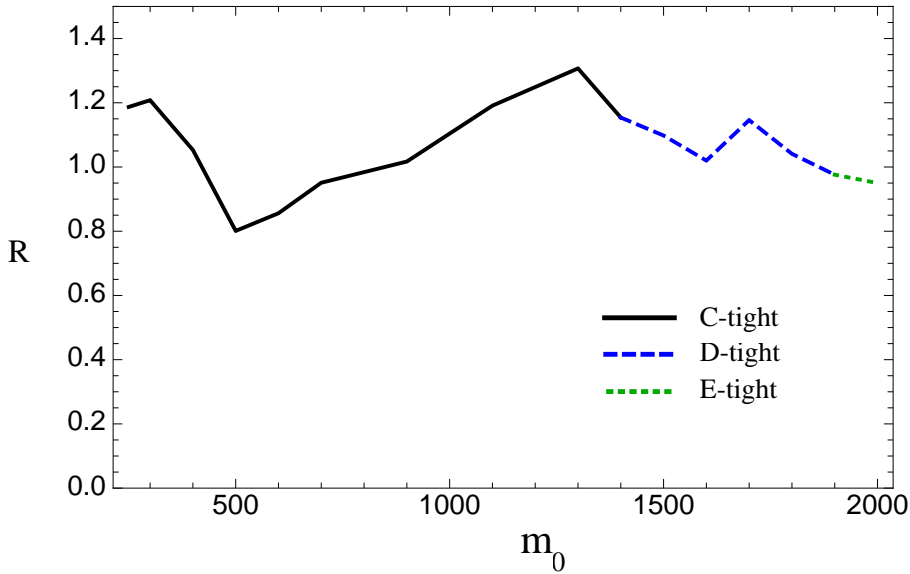, width=130mm}
\end{center}
\vspace*{-8mm}
\caption{The ratio $R$ of the number of signal events $N_{SE}$ from our
simulation within the cMSSM along the 95\% CL exclusion line in
\cite{ATLAS-CONF-2012-109}, divided by the upper limits given in Table~1
from \cite{ATLAS-CONF-2012-109}. The colours indicate the most
constraining search channel: black (full) for C-tight, blue (dashed) for
D-tight and green (dotted) for E-tight.}
\label{fig:1}
\end{figure}

We see that $R$ deviates from 1 by up to $\pm 30\%$, which we take as
uncertainty in the number of signal events after cuts due to our
simulation (it is considerably larger than our statistical error). This
error is added linearly to the error on production cross sections from
scale and PDF choices (which are slightly squark and gluino mass
dependent). For a given value of $m_0$ we determine three different
values of $M_{1/2}$: one such that the number of signal events in the
sNMSSM coincides with the 95\%~CL UL in Table~1 (for the most
constraining channel), and two more such that the number of signal
events in the sNMSSM coincides with the 95\%~CL UL $\pm$ the relative
error obtained as before. This leads to an exclusion curve in the
$m_0-M_{1/2}$ plane, as well as to curves which represent our errors on
the number of signal events.

\section{Results and discussion}

After simulation of a variety of points in the $m_0-M_{1/2}$ plane in
the semi-constrained NMSSM and application of the cuts, we require that
accepted points give a number of signal events below the 95\% CL upper
limit in each search channel, $\pm$ the uncertainty as determined above.
This procedure generates the bounds in the $m_0-M_{1/2}$ plane shown in
Fig.~2, together with the uncertainties indicated by dashed lines. (The
colours in Fig.~2 indicate the most constraining search channels: red for
the jets + $E_T^{\mathrm{miss}}$ channels C-tight, D-tight or E-tight
from \cite{ATLAS-CONF-2012-109}, and blue for the multijet +
$E_T^{\mathrm{miss}}$ channels 9j55 or 8j80 from
\cite{ATLAS-CONF-2012-103}.)

\begin{figure}[h!]
\begin{center}
\epsfig{file=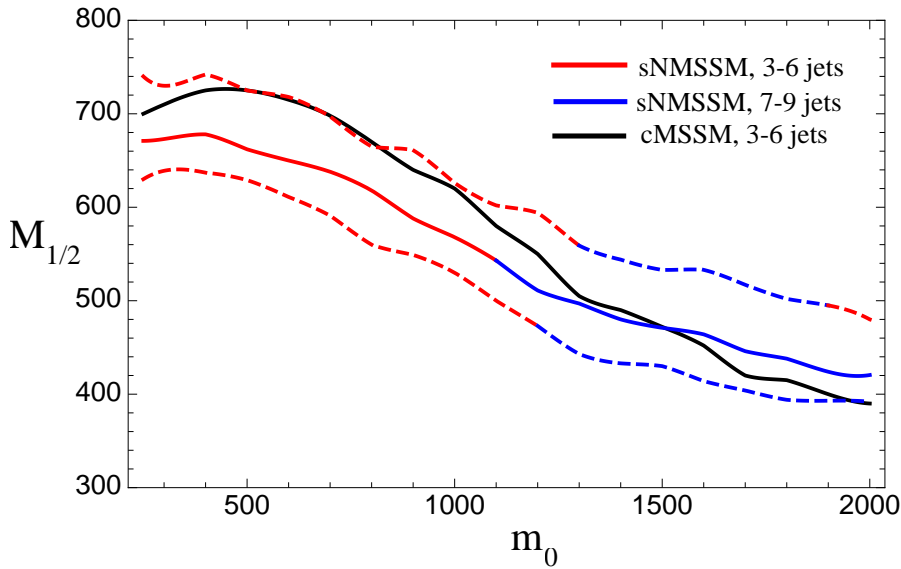, width=135mm}
\end{center}
\vspace*{-8mm}
\caption{Bounds in the $m_0-M_{1/2}$ plane in the semi-constrained
NMSSM (dashed lines indicate our error bars), and ATLAS bounds for the
cMSSM with $\tan\beta=10$ and $A_0=0$ from \cite{ATLAS-CONF-2012-109} in
black. The colours indicate the most constraining search channel for the
NMSSM: Red for the jets + $E_T^{\mathrm{miss}}$ channels C-tight,
D-tight or E-tight from \cite{ATLAS-CONF-2012-109}, and blue for the
multijet + $E_T^{\mathrm{miss}}$ channels 9j55 or 8j80 from
\cite{ATLAS-CONF-2012-103}.}
\label{fig:2}
\end{figure}

Note that, for a given value of $m_0$, the upper and lower error lines
can originate from different search channels. For comparison, we also
show the bounds obtained by ATLAS for the cMSSM with $\tan\beta=10$ and
$A_0=0$ as a black line.

We see that for lower values of $m_0$, the cMSSM bounds are alleviated
due to NMSSM specific sparticle decay cascades. For larger values of
$m_0$ the bounds within the semi-constrained NMSSM seem stronger. This
is due to the fact that in this regime bounds from the multijet channels
\cite{ATLAS-CONF-2012-103} become stronger than the bounds from the
otherwise dominant channel E-tight; notice that the bounds from the multijet
channels \cite{ATLAS-CONF-2012-103} are not included in the
ATLAS bounds for the cMSSM with $\tan\beta=10$ and $A_0=0$ from
\cite{ATLAS-CONF-2012-109}. We have verified that the bounds from the
multijet channels \cite{ATLAS-CONF-2012-103} would also dominate in the
cMSSM for $m_0 \gsim 1400$~GeV, leading to cMSSM bounds somewhat
stronger than those given in \cite{ATLAS-CONF-2012-109}.

\begin{figure}[h!]
\begin{center}
\epsfig{file=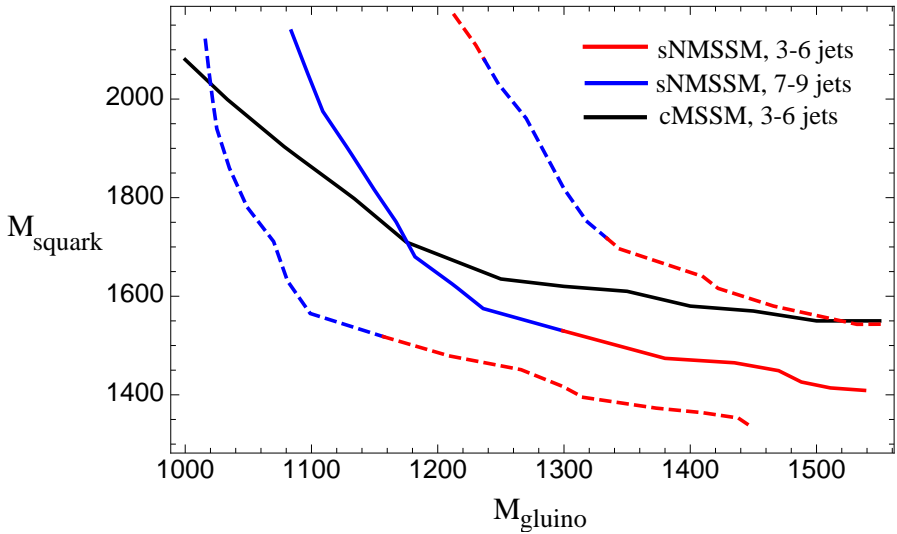, width=125mm}
\end{center}
\vspace*{-8mm}
\caption{Bounds in the $M_{\mathrm{squark}} - M_{\mathrm{gluino}}$ plane
in the semi-constrained NMSSM (dashed lines indicate our error bars),
and ATLAS bounds for the cMSSM with $\tan\beta=10$ and $A_0=0$ from
\cite{ATLAS-CONF-2012-109} in black. Colour/line code as in
Fig.~\ref{fig:2}.}
\label{fig:3}
\end{figure}

The corresponding bounds in the $M_{\mathrm{squark}} -
M_{\mathrm{gluino}}$ plane are shown in Fig.~\ref{fig:3}. Here the
region $m_0 \lsim 1500~$GeV ($M_{1/2} \gsim 450$~GeV), where the bounds
from the jets + $E_T^{\mathrm{miss}}$ channels are somewhat weaker in
the sNMSSM than in the cMSSM, corresponds to $M_{\mathrm{squark}} \lsim
1700$~GeV ($M_{\mathrm{gluino}}\gsim 1200$~GeV).

The different cMSSM and sNMSSM bounds are due to having distinct
sparticle decay cascades. In the following, we discuss these cascades
(which depend on $m_0$) in more detail.

For $m_0 \lsim 1000$~GeV, the sparticle production cross section is
dominated by up/down squark pair production. In the cMSSM with
$\tan\beta=10$ and $A_0=0$, the $\mu$-parameter and hence the higgsino
masses are relatively large ($\approx 800$~GeV), the binos and winos
are approximately eigenstates with masses $\sim 0.4\times M_{1/2}$,
$\sim 0.8\times M_{1/2}$, respectively, and the bino is the LSP
(violating generally WMAP bounds on the relic density). The dominant
decays of the right-handed squarks $\tilde{q}_R$ and
left-handed squarks $\tilde{q}_L$ are
\begin{align}
\tilde q_R \to q + \chi_1^0 \,\, ; \quad 
\tilde q_L \to q + \chi_1^\pm \to q +W^\pm + \chi_1^0 \,,
\end{align}
where $\chi_1^0$ is essentially bino-like and $\chi_1^\pm$ essentially
wino-like. Hence the
dominant decay cascades are relatively short.

In the semi-constrained NMSSM with a SM-like Higgs mass of $\sim
125$~GeV and a dark matter relic density consistent with WMAP
constraints, the effective $\mu$-parameter (and hence the higgsino
masses), as well as the singlino mass parameter $2\kappa s$, are
relatively small, in the $115-250$~GeV range. Apart from alleviating the
``little fine-tuning problem'', such higgsino and singlino mass
parameters generate large mixing angles in the neutralino sector. The
LSP, the lightest neutralino, is a mixture of higgsinos and singlino.
The mostly bino-like neutralino is $\chi_4^0$, i.e. not the LSP, and the
lighter chargino $\chi_1^\pm$ is essentially higgsino-like. Hence the
dominant decays of the right-handed and left-handed squarks are
\begin{align}
& \tilde q_R \to q + \chi_4^0 \to q + W^\mp + \chi_1^\pm
 \to  q + W^\mp + W^{\pm *} + \chi_1^0\,\, ; \\
& \tilde q_L \to q + \chi_2^\pm \to q + Z/H + 
\chi_1^\pm \to  q + Z/H +W^{\pm *} + \chi_1^0\,.
\end{align}
The squark decay cascades lead to considerably more final states in the
NMSSM, implying less missing transverse momentum and less $p_T$ per jet
compared to the cMSSM. This explains the lower number of signal events,
and the somewhat lower bound in the $m_0-M_{1/2}$ plane. An example for
a benchmark point with these properties (with $m_0=600$~GeV and
$M_{1/2}=650$~GeV) is given in Table \ref{tab:bmp}.

\begin{table}[ht!]
\begin{center}
\begin{tabular}{ll}
\begin{minipage}{10.3cm} 
\begin{tabular}{||l|c||l|c||} \hline \hline
%\ \ Parameters & & \phantom{eV}P2\phantom{eV}&text\\ \hline\hline
$\lambda\ (M_\text{SUSY})$ & 0.64  & $M_{H_1}$ &123
\\ \hline
$\kappa\ (M_\text{SUSY}) $ & 0.36  & $M_{H_2}$ &125
\\ \hline
$\tan\beta\ (M_\text{SUSY})$ & 3.02  & $M_{H_3}$ &377
\\ \hline
$\mu_{\text{eff}}\ (M_\text{SUSY})$ &  120 & $R_2^{\gamma\gamma}$ (ggF) &1.73
\\ \hline
$M_{1/2}$ & 650  & $R_2^{ZZ}$ (ggF) &1.05
\\ \hline
$m_0$ & 600  & $M_{\chi_1^0}$ (LSP)&76.5
\\ \hline
$A_0$ & -1262  & $\tilde{H}_d$ comp. of $\chi_1^0$ & 0.49
\\ \hline
$A_\lambda$ & -426  &$\tilde{H}_u$ comp. of $\chi_1^0$ & 0.72
\\ \hline
$A_\kappa$ &  -176 &$\tilde{S}$ comp. of $\chi_1^0$ & 0.43
\\ \hline\hline
$\left<M_{\text{squarks}\,\tilde{u},\tilde{d}}\right>$ &  1425 &
$M_{\chi_2^0}$ &159
\\ \hline
$M_\text{gluino}$ &  1490 & $M_{\chi_3^0}$ &196
\\ \hline
$M_{\tilde{t}_1}$ &  604 &$M_{\chi_4^0}$ (bino) &283
\\ \hline\hline
$\Omega h^2$ & 0.0941 & $M_{\chi_1^\pm}$ (higgsino) &112
\\ \hline
$\sigma^p_{\text SI}$  & $8.4\times 10^{-10}$ &
$M_{\chi_5^0},M_{\chi_2^\pm}$ (winos) &540
\\ \hline\hline
\end{tabular}
\end{minipage}
&
\begin{minipage}{5.1cm} 
\begin{tabular}{||l||c||} \hline \hline
Decays& BR(\%) \\
\hline\hline
$\tilde{u}_L \to \chi_5^0 +u$ &  31
\\ \hline
$\tilde{u}_L \to \chi_2^+ +d$ &  62
\\ \hline
$\tilde{d}_L \to \chi_5^0 +d$ &  32
\\ \hline
$\tilde{d}_L \to \chi_2^- +u$ &  65
\\ \hline
$\tilde{u}_R \to \chi_4^0 +u$ &  94
\\ \hline
$\tilde{d}_R \to \chi_4^0 +d$ &  94
\\ \hline\hline
$\chi_4^0 \to \chi_1^\pm + W^\mp$ & 69
\\ \hline
$\chi_4^0 \to \chi_1^0 + H_1$ & 18
\\ \hline
$\chi_5^0 \to \chi_1^\pm + W^\mp$ & 45
\\ \hline
$\chi_5^0 \to \chi_2^0 + Z$ & 17
\\ \hline
$\chi_5^0 \to \chi_1^0 + H_1$ & 15 
\\ \hline\hline
$\chi_1^\pm \to \chi_1^0 + W^*$ & 100
\\ \hline
$\chi_2^\pm \to \chi_1^\pm + Z$ &  23
\\ \hline
$\chi_2^\pm \to \chi_1^\pm + H_1$ &  20
\\ \hline
$\chi_2^\pm \to \chi_1^0 + W^\pm$ &  21
\\ \hline
$\chi_2^\pm \to \chi_2^0 + W^\pm$ &  17
\\ \hline\hline
\end{tabular}
\end{minipage}
\end{tabular}
\end{center}
\caption{Input parameters, spectrum and some branching fractions of a
benchmark point with $m_0=600$~GeV, $M_{1/2}=650$~GeV. All dimensionful
quantities are given in GeV. $\sigma^p_{\text SI}$ denotes the spin
independent LSP-proton cross section, for which the present XENON100
bound is $\lsim  3\times 10^{-9}$ for a LSP mass of $\simeq 77$~GeV.
$R_2^{\gamma\gamma}$, $R_2^{ZZ}$ denote the signal cross sections of
$H_2$ relative to the SM, which are given in the gluon fusion production
mode (ggF). The reduced signal cross sections of $H_1$ in these channels are
about 0.25. $\tilde{H}_{{u,d}}$ and $\tilde{S}$ denote the higgsino and
the singlino components of $\chi_1^0$, respectively.}
\label{tab:bmp}
\end{table}

For $m_0$ close to $250$~GeV (and $M_{1/2} \lsim 800$~GeV), another phenomenon
appears in the semi-constrained NMSSM with a SM like Higgs mass of $\sim
125$~GeV: The Higgs mass requires a non-universal soft Higgs mass term
$m_{H_u}$, which is considerably larger than $m_0$ at the GUT scale. This
has some impact on the running squark and notably on the slepton masses
from the GUT to the weak scale, leading to light sleptons. In fact, the
LEP2 bound on light slepton masses of $\sim 100$~GeV leads to a lower
bound $m_0 \gsim 250$~GeV on the parameter space. Moreover, the mostly
bino-like neutralino decays dominantly into sleptons in this region. We
have checked that constraints from searches in the channels including
isolated leptons in \cite{ATLAS-CONF-2012-104} are satisfied in
this region, and the dominant constraints still originate from the
channel C-tight.

For $m_0 \gsim 1000$~GeV, the sparticle production cross section becomes
dominated by squark-gluino production and, for $m_0 \gsim 1400$~GeV, by
gluino pair production. In the cMSSM, gluinos undergo three-body decays
involving virtual squarks, with some preference for the somewhat lighter
squarks of the third generation. In the semi-constrained NMSSM, the
non-universal soft Higgs mass term $m_{H_u}$, as well as the larger value
for the top Yukawa coupling $h_t$ (due to the lower value of
$\tan\beta$), lead to lighter stop masses. Apart from alleviating the
``little fine-tuning problem'', such stop masses imply dominant
 gluino two-body decays into the top quark + the
lighter top squark (practically 100\%). 
The latter decays dominantly into a bottom quark +
the lighter chargino. (As discussed above, we verify that present
bounds from top squark searches are satisfied.)
Hence the final states involve a large number of
jets, but a somewhat reduced missing transverse momentum. Whereas many
of these final states would pass the cuts in the searches for jets and
missing transverse momentum in \cite{ATLAS-CONF-2012-109}, the searches
for large jet multiplicities in \cite{ATLAS-CONF-2012-103} become now
more relevant and, in fact, the constraints from the multijet
channel are now dominant. For this reason, the constraints for $m_0 \gsim
1400$~GeV in the semi-constrained NMSSM are stronger than those from the
channels D-tight and E-tight on the cMSSM in \cite{ATLAS-CONF-2012-109}.
This would not be the case if the constraints from the multijet
channel would be applied to the cMSSM. A second benchmark point with these
properties (with $m_0=1900$~GeV and $M_{1/2}=450$~GeV) is given in
Table~\ref{tab:bmp2}.

\begin{table}[ht!]
\begin{center}
\begin{tabular}{ll}
\begin{minipage}{10.3cm} 
\begin{tabular}{||l|c||l|c||} \hline \hline
%\ \ Parameters & & \phantom{eV}P2\phantom{eV}&text\\ \hline\hline
$\lambda\ (M_\text{SUSY})$ & 0.61  & $M_{H_1}$ &95
\\ \hline
$\kappa\ (M_\text{SUSY}) $ & 0.37  & $M_{H_2}$ &126
\\ \hline
$\tan\beta\ (M_\text{SUSY})$ & 2.31  & $M_{H_3}$ &331
\\ \hline
$\mu_{\text{eff}}\ (M_\text{SUSY})$ &  126 & $R_2^{\gamma\gamma}$ (ggF) &1.78
\\ \hline
$M_{1/2}$ & 450  & $R_2^{ZZ}$ (ggF) &1.40
\\ \hline
$m_0$ & 1900  & $M_{\chi_1^0}$ (LSP)&79
\\ \hline
$A_0$ & -875  & $\tilde{H}_d$ comp. of $\chi_1^0$ & 0.54
\\ \hline
$A_\lambda$ & -296  &$\tilde{H}_u$ comp. of $\chi_1^0$ & 0.68
\\ \hline
$A_\kappa$ &  -385 &$\tilde{S}$ comp. of $\chi_1^0$ & 0.32
\\ \hline\hline
$\left<M_{\text{squarks}\,\tilde{u},\tilde{d}}\right>$ &  2070 &
$M_{\chi_2^0}$ &162
\\ \hline
$M_\text{gluino}$ &  1140 & $M_{\chi_3^0}$ &192
\\ \hline
$M_{\tilde{t}_1}$ &  555 &$M_{\chi_4^0}$ (bino) &212
\\ \hline\hline
$\Omega h^2$ & 0.105 & $M_{\chi_1^\pm}$ (higgsino) &109
\\ \hline
$\sigma^p_{\text SI}$  & $7.2\times 10^{-9}$ &
$M_{\chi_5^0},M_{\chi_2^\pm}$ (winos) &398
\\ \hline\hline
\end{tabular}
\end{minipage}
&
\begin{minipage}{5.1cm} 
\begin{tabular}{||l||c||} \hline \hline
Decays& BR(\%) \\
\hline\hline
$\tilde{g} \to \tilde{t}_1 + t$ &  100
\\ \hline
${\tilde{t}_1} \to \chi_1^+ +b$ &  51
\\ \hline
${\tilde{t}_1} \to \chi_1^0 +t$ &  22
\\ \hline
${\tilde{t}_1} \to \chi_2^0 +t$ &  18
\\ \hline\hline
$\chi_2^0 \to \chi_1^0 + Z^*$ & 79
\\ \hline
$\chi_2^0 \to \chi_1^\pm + W^*$ & 20
\\ \hline
$\chi_1^\pm \to \chi_1^0 + W^*$ & 100
\\ \hline\hline
\end{tabular}
\end{minipage}
\end{tabular}
\end{center}
\caption{Input parameters, spectrum and some branching fractions of a benchmark
point with $m_0=1900$~GeV, $M_{1/2}=450$~GeV.  The reduced signal cross sections of
$H_1$ in the channels $bb$, $\gamma\gamma$ and $ZZ$ (in the gluon fusion production
mode) are $\sim 0.025$.}
\label{tab:bmp2}
\end{table}

\bigskip

We conclude that the present bounds on $m_0$ and $M_{1/2}$ in the
semi-constrained NMSSM with a SM like Higgs mass of $\sim 125$~GeV are
somewhat alleviated with respect to the cMSSM for $m_0 \lsim 1000$~GeV
(mostly for $m_0 \sim 500$~GeV) due to the longer and more complicated sparticle
decay cascades. This phenomenon could have been anticipated. Here we
have studied it concretely with the result that the lower bound on
$M_{1/2}$, at fixed $m_0$, decreases by at most 50~GeV for $m_0 \sim
500$~GeV; for $m_0 \gsim
1100$~GeV ($M_{\mathrm{squark}} \gsim 1500$~GeV), the lower bound 
arises from multijet searches, whereas those of the cMSSM 
were derived from jet and missing transverse 
momentum search channels.
%% but that the lower bound on
%% $M_{1/2}$ is due to searches for multijets for $m_0 \gsim
%% 1100$~GeV ($M_{\mathrm{squark}} \gsim 1500$~GeV). 
The central line in
Fig.~2 serves to compare the bounds to the cMSSM whereas, to be
conservative, the lower dashed line including our uncertainties should
be used for bounds in the $m_0-M_{1/2}$ plane in the semi-constrained
NMSSM. Moreover it is likely that, within the general NMSSM (as in the
phenomenological MSSM in \cite{CahillRowley:2012kx}), lower bounds on
sparticle masses are much weaker. We expect that a large variety of
corresponding scenarios are possible within the general NMSSM, which
will require more dedicated studies.

\section*{Acknowledgements}

UE acknowledges partial support from the French ANRs STR-\-COSMO and
LFV-CPV-LHC, and UE and AMT from the European Union FP7 ITN INVISIBLES
(Marie Curie Actions,~PITN-GA-2011-289442). DD thanks the Department
of Theoretical Physics, Indian Association for the Cultivation of
Science (IACS), and Harish-Chandra Research Institute (HRI), Allahabad
for kind hospitality during the final stages of this work.

%\newpage

\end{document}